\newcommand{\CLASSINPUTtoptextmargin}{2cm}
\newcommand{\CLASSINPUTbottomtextmargin}{2cm}
\documentclass[journal, a4paper]{IEEEtran}

\usepackage{graphicx}

\usepackage{caption}
\usepackage{listings}

\ifCLASSOPTIONcompsoc
  \usepackage[nocompress]{cite}
\else
  \usepackage{cite}
\fi

\begin{document}
\bstctlcite{IEEEexample:BSTcontrol}

\title{Application Level Authentication for Ethereum Private Blockchain Atomic Crosschain Transactions}

\author{
    \IEEEauthorblockN{Peter Robinson}
    \IEEEauthorblockA{PegaSys, ConsenSys and University of Queensland\\
    peter.robinson@consensys.net peter.robinson@uqconnect.edu.au}
}

\maketitle

\begin{abstract}
Atomic Crosschain Transaction technology allows composable programming across private Ethereum blockchains. It allows for inter-contract and inter-blockchain function calls that are both synchronous and atomic: if one part fails, the whole call graph of function calls is rolled back. Traditional Ethereum contract functions can limit which accounts can call them by specialised application program logic. This is important as it allows application developers to specify which callers can execute functions that update contract state. In this paper we introduce the strategy required to restrict which contracts on one blockchain can call a function in a contract that is deployed on another blockchain. We show that validating the \texttt{Originating Blockchain Id} (the blockchain the crosschain function call started on), \texttt{From Blockchain Id}, and \texttt{From Account} provides contracts with certainty that a function call came from a specific contract on a specific blockchain.
\end{abstract}

\IEEEpeerreviewmaketitle

\section{Introduction}
Atomic Crosschain Transactions \cite{robinson2019b} for Ethereum Private Sidechains \cite{robinson2018a} and private Ethereum blockchains allow for inter-contract and inter-blockchain function calls that are both synchronous and atomic. Atomic Crosschain Transactions are special nested Ethereum transactions that include additional fields to facilitate the atomic behaviour securely. This new type of Ethereum transaction has great promise, but introduces a new set of challenges.

\begin{figure}[!b]
  \includegraphics[width=\linewidth]{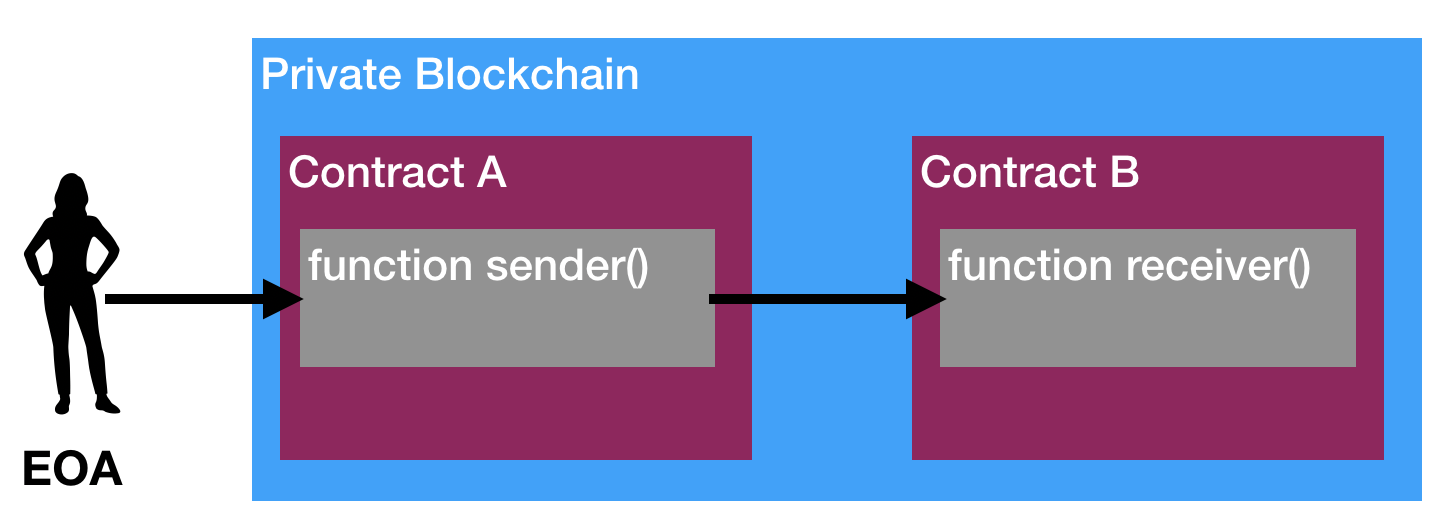}
  \caption{Traditional Transaction Function Call within One Blockchain}
  \label{fig:example-one-blockchain}
\end{figure}

Traditional Ethereum transactions \cite{wood2016a} execute within a single blockchain. For example, in Fig.~\ref{fig:example-one-blockchain} an Externally Owned Account (EOA) submits a transaction that calls the function \texttt{sender} in \texttt{Contract A} that in turn calls the function \texttt{receiver} in \texttt{Contract B}. \texttt{Contract B} could be a simple contract that holds data and has little or no business logic. \texttt{Contract A} may be a complex contract that holds the majority or all of the business logic of the application. The business logic may need to change over time. Additionally, like any complex software, a complex contract may have defects which need to be resolved. The typical approach to this scenario in a blockchain setting is to deploy a new version of \texttt{Contract A} and have \texttt{Contract B}'s \texttt{receiver} function only allow calls from the newly deployed \texttt{Contract A}\cite{robinson2018solidity}. The Solidity code that would allow this to occur is shown in Listing~\ref{listing:example-one-blockchain}.

On line 2 of the listing \texttt{msg.sender}, the address of the contract or EOA that called this function, is compared against a variable \texttt{authorisedAddress}. This value is the address of the deployed instance of \texttt{Contract A} that is authorised to call the \texttt{receiver} function. The transaction executing the function call is aborted if the two values do not match. This line of code ensures \texttt{Contract B}'s \texttt{receiver} function can only be called by functions in an authorised deployed instance of \texttt{Contract A}.

\begin{lstlisting}[
%  frame=single,
  basicstyle=\footnotesize\ttfamily,
  numbers=left,
stepnumber=1, 
  firstnumber=1,
  numberfirstline=true,
  numbersep=5pt,    
  xleftmargin=0.5cm,
  morekeywords={function, external, require, sender, call, msg},
  label=listing:example-one-blockchain,
  caption=Application Authentication
]
function receiver() external {
  require(msg.sender == authorisedAddress);
  ...
}
\end{lstlisting}

\begin{figure}[!b]
  \includegraphics[width=\linewidth]{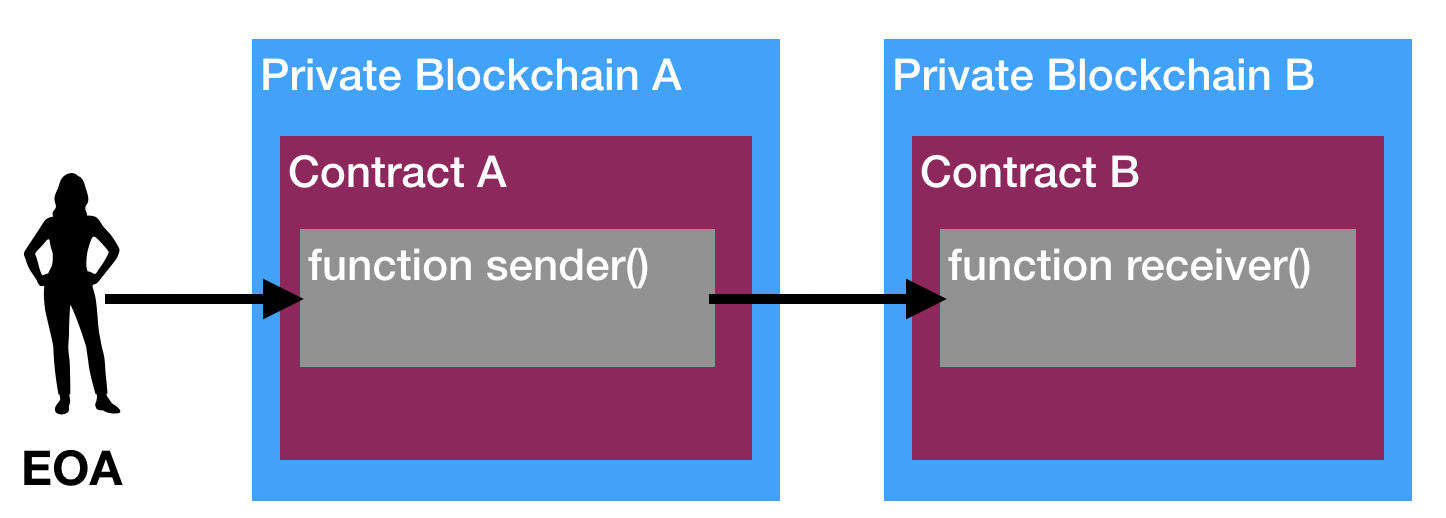}
  \caption{Crosschain Transaction Function Call across Two Blockchains}
  \label{fig:example-two-blockchains}
\end{figure}

The scenario for a crosschain transaction is more complex. In Fig.~\ref{fig:example-two-blockchains} \texttt{Contract A} has been deployed to \texttt{Private Blockchain A} and \texttt{Contract B} has been deployed to \texttt{Private Blockchain B}. An EOA submits a transaction that calls the function \texttt{sender} in \texttt{Contract A} on \texttt{Private Blockchain A} that in turn calls the function \texttt{receiver} in \texttt{Contract B} on \texttt{Private Blockchain B}. This paper describes the application logic required to limit function calls to the function \texttt{receiver} in \texttt{Contract B} to only those coming from \texttt{Contract A} on \texttt{Private Blockchain A}.

\section{Atomic Crosschain Transactions}
\subsection{Nested Transactions}
\begin{figure}
  \includegraphics[width=\linewidth]{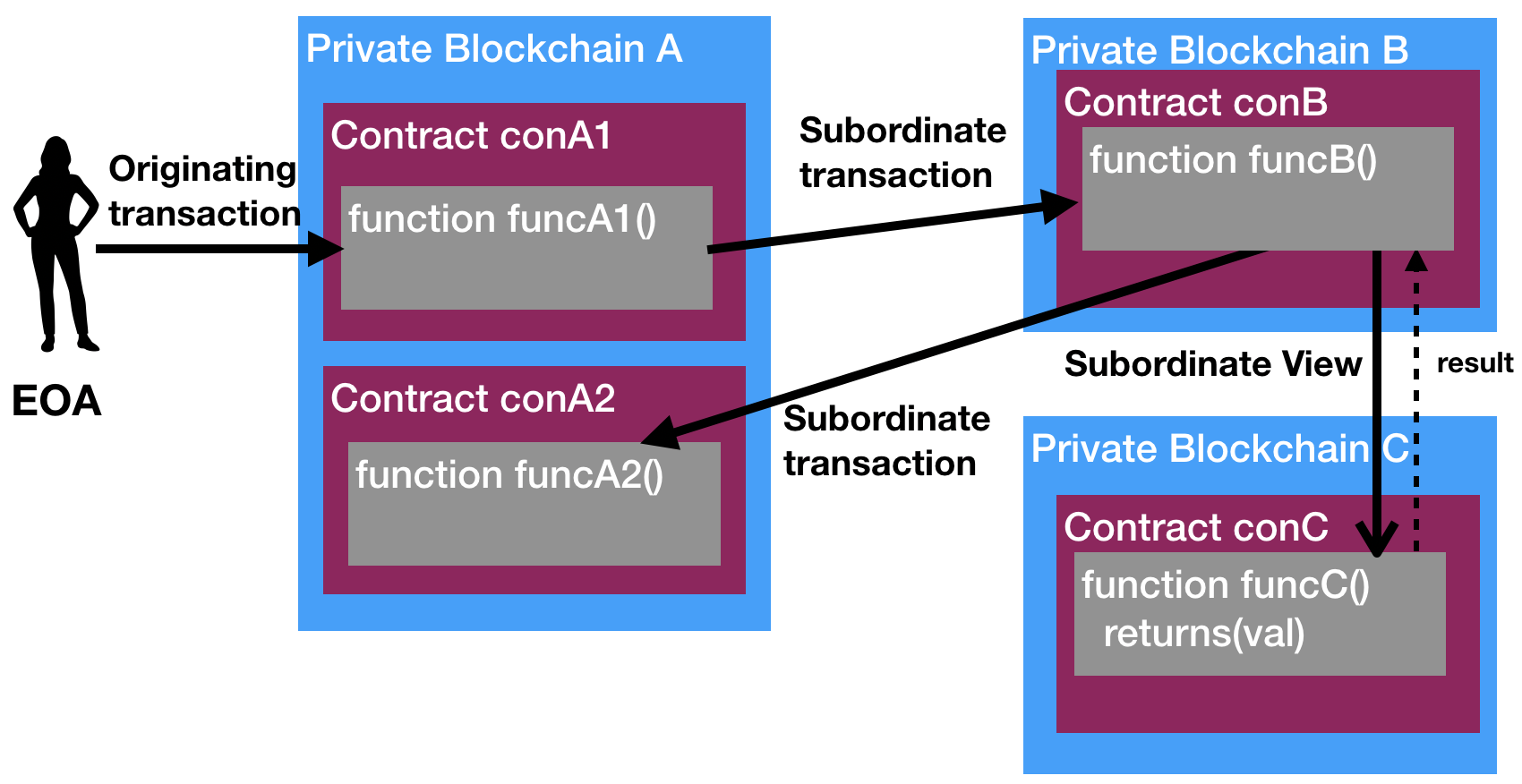}
  \caption{Originating Transaction containing Two Nested Subordinate Transactions and a Subordinate View}
  \label{fig:nested1}
\end{figure}

Atomic Crosschain Transactions are nested Ethereum transactions and views. Transactions are function calls that update state. Views are function calls that return a value but do not update state. Fig.~\ref{fig:nested1} shows a EOA calling a function \texttt{funcA1} in contract \texttt{conA1} on blockchain \texttt{Private Blockchain A}. This function in turn calls function \texttt{funcB}, that in turn calls functions \texttt{funcC} and \texttt{funcA2}, each on separate blockchains. The transaction submitted by the EOA is called the \textit{Originating Transaction}. The transactions that the Originating Transaction causes to be submitted are called Subordinate Transactions. Subordinate Views may also be triggered. In Fig.~\ref{fig:nested1}, a Subordinate View is used to call \texttt{funcC}. This function returns a value to \texttt{funcB}.

\begin{figure}
  \includegraphics[width=\linewidth]{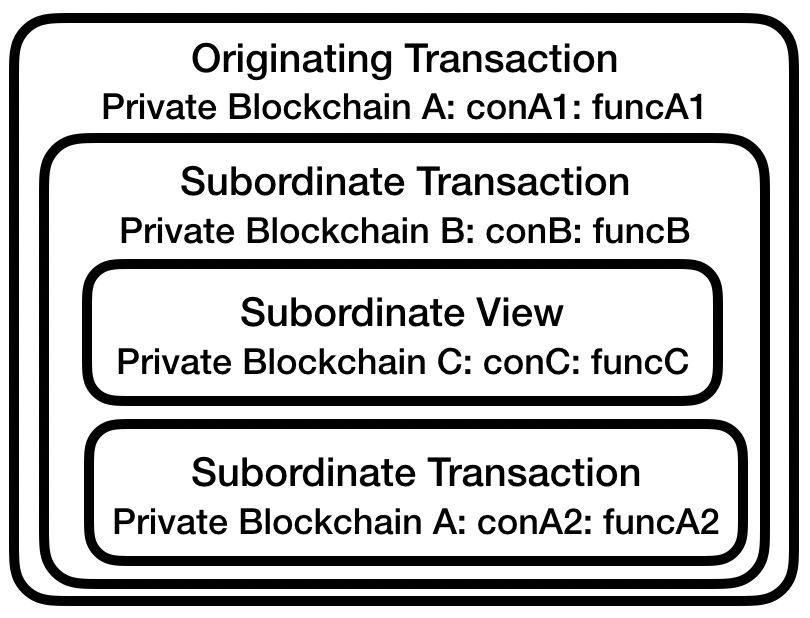}
  \caption{Nested Transactions and Views}
  \label{fig:nested2}
\end{figure}

Fig.~\ref{fig:nested2} shows the nested structure of the Atomic Crosschain Transaction. The EOA user first creates the signed Subordinate View for \texttt{Private Blockchain C}, contract \texttt{conC}, function \texttt{funcC} and the signed Subordinate Transaction for \texttt{Private Blockchain A}, contract \texttt{conA2}, function \texttt{funcA2}. They then create the signed Subordinate Transaction for \texttt{Private Blockchain B}, contract \texttt{conB}, function \texttt{funcB}, and include the signed Subordinate Transaction and View. Finally, they sign the Originating Transaction for \texttt{Private Blockchain A}, contract \texttt{conA1}, function \texttt{funcA1}, including the signed Subordinate Transactions and View.  

When the EOA submits the Originating Transaction to a node, the node processes the transaction using the algorithm shown in Listing~\ref{listing:processing}. If the transaction includes any Subordinate Views, they are dispatched and their results are cached (Lines 1 to 3). The function is then executed (Lines 4 to 17). If a Subordinate Transaction function call is encountered, the node checks that the parameter values passed to the Subordinate Transaction function call match the parameter values in the signed Subordinate Transaction (Lines 6 to 8). If a Subordinate View function call is encountered, the node checks that the parameters passed to the Subordinate View function call match the parameter values in the signed Subordinate View (Lines 9 and 10). The cached values of the results of the Subordinate View function calls are then returned to the executing code (Line 11). If the execution has completed without error, then each of the signed Subordinate Transactions is submitted to a node on the appropriate blockchain (Nodes 18 to 20).

\begin{lstlisting}[
%  frame=single,
  basicstyle=\footnotesize\ttfamily,
  numbers=left,
stepnumber=1, 
  firstnumber=1,
  numberfirstline=true,
  numbersep=5pt,    
  xleftmargin=0.5cm,
  morekeywords={msg},
  label=listing:processing,
  caption=Originating or Subordinate Transaction Processing
]
For All Subordinate Views {
  Dispatch Subordinate Views & cache results
}
Trial Execution of Function Call {
  While Executing Code {
    If Subordinate Transaction function called {
      check expected & actual parameters match.
    } 
    Else If Subordinate View function is called {
      check expected & actual parameters match
      return cached results to code
    } 
    Else {
      Execute Code As Usual
    }
  }
}
For All Subordinate Transactions {
  Submit Subordinate Transactions
}
\end{lstlisting}

\subsection{Blockchain Signing and Threshold Signatures}
BLS Threshold Signatures \cite{bls-threshold, bls-threshold-youtube} combines the ideas of threshold cryptography \cite{shamir1979} with Boneh-Lynn-Shacham(BLS) signatures \cite{bls2004}, and uses a Pedersen commitment scheme \cite{ped1991} to ensure verifiable secret sharing. The scheme allows any \texttt{M} validator nodes of the total \texttt{N} validator nodes on a blockchain to sign messages in a distributed way such that the private key shares do not need to be assembled to create a signature. Each validator node creates a signature share by signing the message using their private key share. Any \texttt{M} of the total \texttt{N} signature shares can be combined to create a valid signature. Importantly, the signature contains no information about which nodes signed, or what the threshold number of signatures (\texttt{M}) needed to create the signature is.

The Atomic Crosschain Transaction system uses BLS Threshold Signatures to prove that information came from a specific blockchain. For example, in Fig.~\ref{fig:nested1}, nodes on \texttt{Private Blockchain B} can be certain of results returned by a node on \texttt{Private Blockchain C} for the function call to \texttt{funcC}, as the results are threshold signed by the validator nodes on \texttt{Private Blockchain C}. Similarly, validator nodes on \texttt{Private Blockchain A} can be certain that validator nodes on \texttt{Private Blockchain B} have mined the Subordinate Transaction, locked contract \texttt{conB} and are holding the updated state as a provisional update because validator nodes sign a \textit{Subordinate Transaction Ready} message indicating that the Subordinate Transaction is ready to be committed.

\subsection{Crosschain Coordination}
\textit{Crosschain Coordination Contracts} exist on \textit{Coordination Blockchains}. They allow validator nodes to determine whether the provisional state updates related to the Originating Transaction and Subordinate Transactions should be committed or discarded. The contract is also used to determine a common time-out for all blockchains, and as a repository of Blockchain Public Keys. 

When a user creates a Crosschain Transaction, they specify the Coordination Blockchain and Crosschain Coordination Contract to be used for the transaction, and the time-out for the transaction in terms of a block number on the Coordination Blockchain. The validator node that they submit the Originating Transaction to (the \textit{Originating Node}) works with other validator nodes on the blockchain to sign a \textit{Crosschain Transaction Start} message. This message is submitted to the Crosschain Coordination Contract to indicate to all nodes on all blockchains that the Crosschain Transaction has commenced. 

When the Originating Node has received Subordinate Transaction Ready messages for all Subordinate Transactions, it works with other validator nodes to create a \textit{Crosschain Transaction Commit} message. This message is submitted to the Crosschain Coordination Contract to indicate to all nodes on all blockchains that the Crosschain Transaction has completed and all provisional updates should be committed. If an error is detected, then a \textit{Crosschain Transaction Ignore} message is created and submitted to the Crosschain Coordination Contract to indicate to all nodes on all blockchains that the Crosschain Transaction has failed and all provisional updates should be discarded. Similarly, if the transaction times-out, all provisional updates will be discarded.

\subsection{Crosschain Transaction Fields}
Originating Transactions, Subordinate Transactions, and Subordinate Views contain the fields shown in Table~\ref{table:fields}. Some of the information in the standard Ethereum transaction fields are exposed to blockchain application contract code, such as the \texttt{value} field via the Solidity code \texttt{msg.value}. The new extended crosschain transaction fields are made available to blockchain application contract code via a precompile contract.

\begin{table}
  \centering
    \begin{tabular}{| l | l |}
    \hline
    Field & Description  \\
       \hline
       \hline
        \multicolumn{2}{|l|}{Standard Ethereum Transaction Fields} \\
       \hline
Nonce     & Per-account, per-blockchain transaction number. \\
       \hline
GasPrice & Amount offered to pay for gas for the transaction.\\
       \hline
GasLimit  &  Maximum gas which can be used by the transaction.\\
       \hline
To            & Address of the account to send the value to, or the  \\
                & address of a contract to call.\\
       \hline
Value       & Amount of Ether to transfer.\\
       \hline
Data         & Encoded function signature and parameter values.\\
       \hline
V              & Part of the transaction digital signature \& blockchain \\
                & identifier this transaction must execute on.\\
       \hline
R              & Part of the transaction digital signature.\\
       \hline
S              & Part of the transaction digital signature.\\
       \hline
       \hline
        \multicolumn{2}{|l|}{Additional Crosschain Transaction Fields} \\
       \hline
    Type     & Type of crosschain transaction (e.g. Originating \\
                 & Transaction) \\
    \hline
Coordination & Blockchain identifier of Coordination Blockchain to  \\
Blockchain Id & use for this transaction.\\
       \hline
Crosschain  & Address of the Crosschain Coordination Contract \\
Coordination &  to use for this transaction. \\
Contract       & \\
       \hline
Crosschain   & Coordination Blockchain block number when this\\ 
Transaction  & transaction will time out. \\
Time-out      & \\
       \hline
Crosschain  &  Identifies this crosschain transaction. \\
Transaction Id & \\
       \hline
Originating  & Blockchain identifier of the blockchain the  \\
Blockchain Id &  Originating Node is on. \\
       \hline
From             & Blockchain identifier of the blockchain that the   \\
Blockchain Id &  function call executed on that resulted in this \\
                       & Subordinate Transaction or View being submitted. \\
       \hline
From            & \textit{To} address from the transaction or view that resulted \\
Address        & in this Subordinate  Transaction or View.\\
       \hline
Subordinates & List of Subordinate Transactions and Subordinate  \\
                       & Views that are called directly from this transaction \\
                       & or view. \\
       \hline
  \end{tabular}
  \caption{Crosschain Transaction Fields}
  \label{table:fields}
\end{table}

All nodes that process the transaction check that the \texttt{Coordination Blockchain Id}, \texttt{Crosschain Co- ordination Contract}, \texttt{Crosschain Transaction Time-out}, \texttt{Crosschain Transaction Id}, and \texttt{Originating Blockchain Id} are consistent across the transaction or view they are processing, and the nested Subordinate Transactions and Views. The nodes also check that the \texttt{To} address and \texttt{From Address}, and the blockchain identifier obtained from the \texttt{V} field and the \texttt{From Blockchain Id} match across transactions and views.

The \texttt{To} address is the address of the contract containing the function called on a blockchain. For example, the function (f1) in contract (c1) could call a function (f2) in another contract (c2) on the same blockchain (b1). The second contract (c2) could call a function (f3) in a contract (c3) on another blockchain (b2) via a Subordinate Transaction. The \texttt{From Address} of the Subordinate Transaction will match the \texttt{To} address of the transaction on the first blockchain (b1). This will be the address first contract (c1). It will however, not match the address of the second contract (c2), which is the function that caused the Subordinate Transaction to be triggered.

\section{Application Authentication}
As with traditional Ethereum transactions, the type of application level authentication required for a Crosschain Transaction will be application dependent. 
\subsection{No Authentication}
Many functions will need no authentication at all. That is, functions can be designed such that it is safe to execute a transaction or return results of a view to any caller who is able to access the function. 

\subsection{Using \texttt{msg.sender} or \texttt{tx.origin}}
From the perspective of each Originating Transaction, Subordinate Transaction or View, \texttt{msg.sender} and \texttt{tx.origin} operate in the same way as a standard Ethereum transaction. That is, if an EOA submitted a transaction that called a function in contract A that then called a function in contract B on the same blockchain, \texttt{msg.sender} for contract B is contract A, and is the EOA for contract A. In both cases \texttt{tx.origin} would be the EOA. In the context of a node processing an Originating Transaction, Subordinate Transaction or View, for the purposes of  \texttt{msg.sender} and \texttt{tx.origin}, the transaction or view appears as a separately signed transaction. Given the similarities with standard Ethereum, \texttt{msg.sender} and \texttt{tx.origin} could be used in the same way as standard Ethereum to authenticate which EOA or contract on the same blockchain called a function call using code similar to that shown in Listing~\ref{listing:example-one-blockchain}. 

A key difference between standard Ethereum views and Subordinate Views is that Subordinate Views are signed. As such, the variables \texttt{msg.sender} and \texttt{tx.origin} can be used within Subordinate Views, whereas they are not set in the context of normal Ethereum views (except for the case of \texttt{msg.sender} when one contract calls another contract). 

\subsection{From Blockchain Id, From Address, and Originating Blockchain Id}
If a contract needs to only respond to calls from a certain contract on a certain blockchain, then the code in Listing~\ref{listing:example-auth} should be used. The code checks that the \texttt{From Blockchain Id} and \texttt{From Address} match the authorised blockchain and address, and checks that the blockchains represented by \texttt{From Blockchain Id} and \texttt{Originating Blockchain Id} are semi-trusted. By semi-trusted it is meant that fewer than \texttt{M} validators operating the blockchain are Byzantine. Note that this scenario implies the contract should allow for any \texttt{msg.sender} and \texttt{tx.origin}. 
\begin{lstlisting}[
%  frame=single,
  basicstyle=\footnotesize\ttfamily,
  numbers=left,
stepnumber=1, 
  firstnumber=1,
  numberfirstline=true,
  numbersep=5pt,    
  xleftmargin=0.5cm,
  morekeywords={address, function, uint256, require, sender, call, msg},
  label=listing:example-auth,
  caption=Crosschain Application Authentication
]
function receiver() external {
  address fromAddr = infoPrecompile(FROM_ADDR);
  uint256 fromBcId = infoPrecompile(FROM_BCID);
  uint256 origBcId = infoPrecompile(ORIG_BCID);
  require(fromAddr == authorisedFromAddress);
  require(fromBcId == authorisedFromBcId);
  require(fromAddr == authorisedOrigBcId);
  ...
}
\end{lstlisting}

\section{Analysis}
\begin{figure}[t]
  \includegraphics[width=\linewidth]{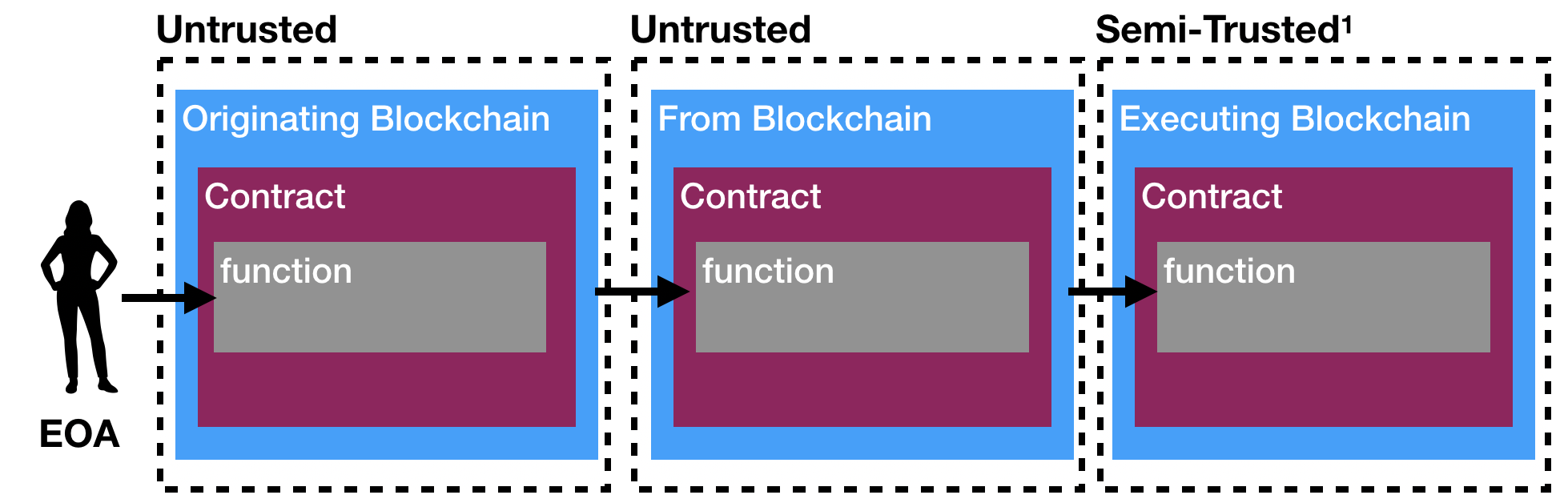}
  \caption{Scenario 1: From and Originating Blockchain Untrusted}
  \label{fig:scenario1}
\end{figure}
\begin{figure}[t]
  \includegraphics[width=\linewidth]{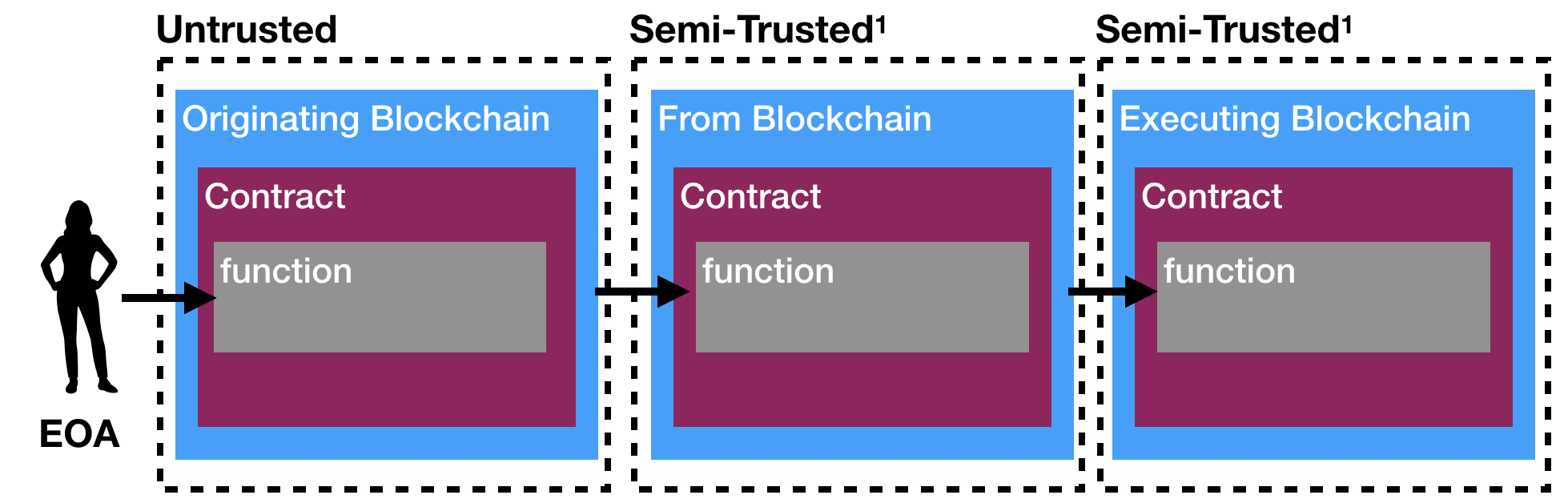}
  \caption{Scenario 2: Originating Blockchain Untrusted}
  \label{fig:scenario2}
\end{figure}
\begin{figure}[t]
  \includegraphics[width=\linewidth]{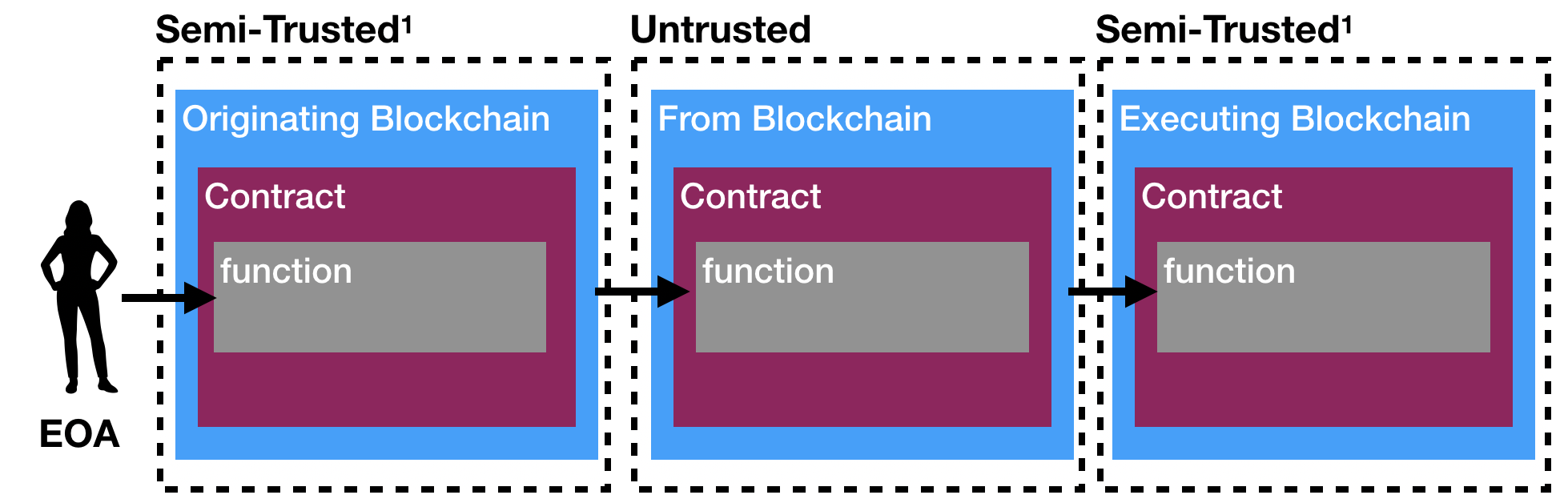}
  \caption{Scenario 3: From Blockchain Untrusted}
  \label{fig:scenario3}
\end{figure}
\begin{figure}[t]
  \includegraphics[width=\linewidth]{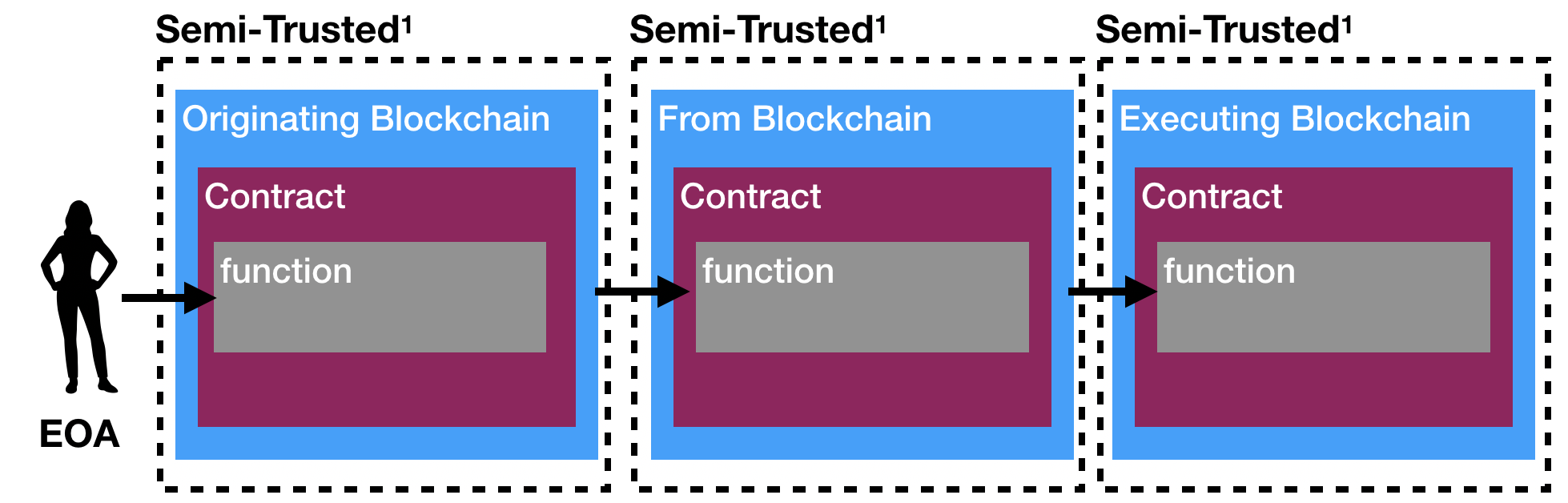}
  \caption[TODO]{Scenario 4: From and Originating Blockchain Semi-Trusted\footnotemark}
  \label{fig:scenario4}
\end{figure}
\footnotetext{Semi-trusted is defined as having fewer than \texttt{M} Byzantine validator node on a blockchain.}
This section analyses the appropriateness of using the \texttt{Originating Blockchain Id}, \texttt{From Blockchain Id}, and \texttt{From Address} fields as a method of authentication, and requiring the blockchains identified by \texttt{Originating Blockchain Id} and \texttt{From Blockchain Id} be semi-trusted. Figures~\ref{fig:scenario1} to \ref{fig:scenario4} show four possible scenarios. The participant could not trust the blockchains represented by \texttt{Originating Blockchain Id}, \texttt{From Blockchain Id} (scenario 1, Figure~\ref{fig:scenario1}), just semi-trust the \texttt{From Blockchain Id} or \texttt{Originating Blockchain Id} (scenario 2, Figure~\ref{fig:scenario2} and scenario 3, Figure~\ref{fig:scenario3}), or semi-trust both blockchains (scenario 4, Figure~\ref{fig:scenario4}). In the figures, the \textit{Originating Blockchain} is the blockchain identified by the \texttt{Originating Blockchain Id}, the \textit{From Blockchain} is the blockchain identified by the \texttt{From Blockchain Id}, and the \textit{Executing Blockchain} is the blockchain executing the transaction that contains the application level authentication logic.

\subsection{Scenario 1}
If neither the Originating  Blockchain or the From Blockchain are trusted, then a nefarious actor operating the validator nodes on the blockchains could maliciously construct a Subordinate Transaction and submit it to the Executing Blockchain. The nefarious actor could create valid Crosschain Transaction Start and Commit messages and submit them to the Coordinating Blockchain, thus making it appear that all nodes on all blockchains should commit all parts of the Crosschain Transaction. 

The nodes on the Executing Blockchain have no basis to trust information in the Subordinate Transaction submitted to them or the Crosschain Transaction status indicated by the Crosschain Coordination Contract. As such, there is no method of application level authentication to restrict which contract on which blockchain can call a function in a contract if neither the Originating  Blockchain nor the From Blockchain are semi-trusted.

\subsection{Scenario 2}
If the From Blockchain is semi-trusted, but the Originating  Blockchain is not trusted, then a nefarious actor could create a malicious Crosschain Transaction. Rather than submitting a Subordinate Transaction to the From Blockchain, they could bypass the blockchain, constructing a malicious Subordinate Transaction with forged From Blockchain Id and From Address, and submit it to the Executing Blockchain. The nefarious actor could create valid Crosschain Transaction Start and Commit messages and submit them to the Coordinating Blockchain, thus making it appear that all nodes on all blockchains should commit all parts of the Crosschain Transaction. 

In this scenario, the nodes on the Executing Blockchain have no way to be certain that the Subordinate Transaction submitted to them originated from the From Blockchain. As such, there is no method of application level authentication to restrict which contract on which blockchain can call a function in a contract if the Originating  Blockchain is not semi-trusted.

\subsection{Scenario 3}
\label{section:scenario3}
If the Originating Blockchain is semi-trusted, but the From Blockchain is not trusted, then a nefarious actor could claim a Subordinate Transaction being executed by the From Blockchain was ready to be committed when it was not. The nefarious actor would not be able to forge the \texttt{From Address} or the \texttt{From Blockchain Id} of the subordinate transaction as validators on the Originating Blockchain would detect the mis-matched \texttt{To} and \texttt{From Address} addresses or blockchains identifiers, and reject the invalid Crosschain Transaction. In this case, they would refuse to mine the Originating Transaction and refuse to create the Crosschain Transaction Start message.

In this scenario, the nodes on the Executing Blockchain are certain that the Subordinate Transaction submitted to them has authentic \texttt{From Address} and \texttt{From Blockchain Id} information. However, there is no certainty that the Subordinate Transaction submitted to the From Blockchain will be committed to that blockchain.

\subsection{Scenario 4}
If both the Originating Blockchain and the From Blockchain are semi-trusted, then a nefarious actor is unable to subvert the protocol. Similarly to section \ref{section:scenario3}, invalid transactions they submit will be rejected by validators nodes on the Originating Blockchain. Validator nodes on the Execution Blockchain can be sure that if the Crosschain Coordination Contract indicates that the transaction should be committed, then all nodes, including the From Blockchain, are ready to commit their provisional updates.

\section{Implementation}
Examples of the Atomic Crosschain Transaction application authentication code is available on github.com\cite{crosschain-github}.

\section{Discussion}
The system assumes that blockchains involved in a crosschain transaction are semi-trusted, where semi-trusted is defined as having fewer than \texttt{M} Byzantine validators nodes operating a blockchain. This assumption of having a threshold number of Byzantine validator nodes is the same type of assumption that Byzantine Fault Tolerant (BFT) consensus protocols make \cite{ibft, ibft2}. As blockchains that support Atomic Crosschain Transaction technology are likely to use a BFT consensus protocol, if more than a threshold number of validator nodes were Byzantine, then the blockchain's consensus protocol, as well as the crosschain transaction protocol, would fail.

\section{Conclusion}
Application programmers need to restrict which callers can call functions in their contracts to update state. Traditional Ethereum security practices are not sufficient for a crosschain transaction context. This paper presents the fundamental building blocks of a crosschain authentication framework on which application-level authentication can be built. In particular, when using Atomic Crosschain Transactions, the \texttt{Originating Blockchain Id}, \texttt{From Blockchain Id}, and the \texttt{From Account} crosschain transaction fields can be used to ensure a function in a contract on a blockchain is only callable from certain contracts on certain blockchains, assuming that the participant that configured the contract trusts that fewer than a threshold number of validators on the blockchains indicated by the \texttt{Originating Blockchain Id}, \texttt{From Blockchain Id} are Byzantine.

\ifCLASSOPTIONcompsoc
  \section*{Acknowledgments}
\else
  \section*{Acknowledgment}
\fi
This research has been undertaken whilst I have been employed full-time at ConsenSys. I acknowledge the support of University of Queensland where I am completing my PhD, and in particular the support of my PhD supervisor Dr Marius Portmann.

I acknowledge the co-authors of the original Atomic Crosschain Transaction paper \cite{robinson2019b} Dr David Hyland-Wood, Roberto Saltini, Dr Sandra Johnson, and John Brainard for their help creating the technology upon which this paper is based. I acknowledge Dr Raghavendra Ramesh for his ideas on improvements to the \textit{From Address} field. I thank Dr Catherine Jones, Dr David Hyland-Wood, Dr Raghavendra Ramesh, Dr Sandra Johnson, John Brainard and Horacio Mijail Anton Quiles for reviewing this paper and providing astute feedback.

\bibliographystyle{IEEEtran}
\bibliography{IEEEabrv,ref}

\end{document}